\documentclass[epj]{svjour}
\usepackage{graphics}
\begin{document}
\title{The Volatility in a Multi-share Financial Market Model}
\author{Adam Ponzi\inst{1,2} 
}                     
%
%
\institute{Department of Physics, Trinity College, Dublin, Ireland. email: adam@phy.tcd.ie \and Hibernian Investment Management, Dublin, Ireland.}
\date{Received: date / Revised version: date}
%
\abstract{Single index financial market models cannot account for the
empirically observed complex interactions between shares in a
market. We describe a multi-share financial market model and compare
characteristics of the volatility, that is the standard deviation of
the price fluctuations, with empirical characteristics. In particular
we find its probability distribution is similar to a log normal
distribution but with a long power-law tail for the large
fluctuations, and that the time development shows superdiffusion. Both
these results are in good quantitative agreement with observations.
} 
\maketitle
\section{Introduction}
\label{intro}

Recent works have shown that financial markets cannot be completely
described by single index models since they do not account for complex
interactions among stocks. Share and currency cross correlation
matrices show some large eigenvalues where the market and certain
groups of companies/currencies move together, against a background
that would be expected from Random Matrix Theory\cite{stan5}. The
largest eigenvalue shows sudden increase at crashes\cite{drozdz}
showing a strong global behaviour at such times. Share
cross-correlations are generally positive\cite{kull}. Recent
results\cite{mant1} show more complex properties of the simultaneous
distribution of individual share returns coupled in a market. These
results indicate that individual stocks in a market are coupled in a
complex fashion and single index models cannot account for such
characteristics.

In a recent paper\cite{physA} we have described a multi-share model of
a financial market and shown that it can account for some inter-share
characteristics as well as for some of the now `well-known'
properties, such as the observation that real market returns
distributions show `fat-tails'. That is, the returns distributions are
Levy stable distributions\cite{LP1} for the central part often
characterised by a parameter $\alpha \simeq 1.4-1.7$, while the wings
(after about 4 standard deviations) fall-off faster as a power-law
(exponent $2 \sim 5$) or possibly as a streatched
exponential\cite{stan2,MS1,stan1,mand1,M1}.

In this paper we will show by numerical simulations that this model
can also account for the empirically observed properties of the {\it
volatility}. Volatility is the local standard deviation of a price
returns time series. The volatility distribution shows an approximate
log-normal distribution for the central part but with tails that seem
to follow a power law with exponent $\simeq 4$,\cite{stan3}.
Furthermore volatility is known to be `clustered'. That is we observe
bursts of relatively high volatility separated by longer periods of
relatively low volatility. This intermittent behaviour is
characterised by a superdiffusion with exponent $\simeq
0.7$,\cite{stan1,M1}.

\section{Model}
\label{sec:1}
Here we describe the model first presented in \cite{tohwa}. A more
detailed discussion of its origins can be found in
\cite{physA}. Precursor models can be found in
\cite{cha1,cha2,badhonnef}.

In this model there are $N$ interacting stocks $i=1,...,N$ where for
example $i=1$ refers to `IBM', $i=2$ refers to `Disney' etc. If we
were considering the S\&P500 then $N\approx 500$. Each stock is
completely characterized by two variables:

1) Excess demand/supply, $s_{i}(t)$. This is a spin variable
$s_{i}(t)=\pm1$ which describes stock $i$'s current market
demand/supply state. That is if $s_{i}(t)=1$ then stock $i$ is in the
excess demand state at time $t$, but if $s_{i}(t)=-1$ then it is in
the excess supply state. In reality excess demand/supply has {\it
magnitude} as well as {\it sign} but in this extreme simplification
only its sign is taken into account. This is because as explained
below we are only interested in modeling the self-reinforcing
persistency characteristics of each stock $i$.  (This model can be
easily generalized by including a excess supply/demand magnitude
$g_{i}(t)$, by, say, randomly choosing the magnitude $g_{i}(t)$ from a
Gaussian (or otherwise) distribution each $t$.)  The price return
$\Delta p_{i}(t)$ for stock $i$ at time $t$ is given by $\Delta
p_{i}(t)=Ns_{i}(t)/2$. The `index' price return $\Delta p(t)$ is then
given by $\Delta p(t)=S(t)/2$ where
$S(t)=\sum_{i=1}^{N}s_{i}(t)$. This is just the usual linear
relationship $dp/dt=Dem(t)-Sup(t)$, where $Dem(t)$ and $Sup(t)$ refer
to the demand and supply at time $t$ respectively. (See for example
\cite{Farmer}).

2) `Speculator Confidence', $V_{i}(t)$ in stock $i$ at time $t$. This
   is a real number which describes the overall (mean) speculator
   expectation of the current excess demand/supply state $s_{i}(t)$
   continuing to the next time step, or reversing. In this model
   speculators are {\it homogenous} in their views.  High confidence
   in a stock means speculators in general expect the current
   demand/supply state $s_{i}(t)$ of that stock to continue, low
   confidence means they expect a {\it reversal}. Our confidence
   $V_{i}(t)$ is cumulative and should be thought of like a `fitness'
   in ecodynamics. It is not an intrinsic property of a stock, but is
   defined only by the current speculators, as a result of their
   expectations, and of course changes with time.

There is no spatial dimension and stocks interact only through the
market macrostate defined by 1) the market excess supply/demand
$G(t)=S(t)/N$ where $G(t)$ is the mean-spin
$G(t)=1/N\sum_{i=1}^{N}s_{i}(t)$, and 2) the {\it market confidence}
(market fitness) defined by $V(t)=1/N\sum_{i=1}^{N}V_{i}(t)$. In real
markets both of these are known (imperfectly) to the speculators. (In
the case of the mean-confidence $V(t)$, there may just be a general
`mood' traders can sense.)

For the dynamics every time step $t$ we calculate $G(t)$ and $V(t)$
and then calculate the {\it relative confidence} $u_{i}(t)$ of stock
$i$ given by $u_{i}(t)=V_{i}(t)-V(t)$. In this model speculators'
expectations are on average completely {\it self-fulfilling} and with
probability $Q_{i}(t)=Q(u_{i}(t))$ where $Q(x)$ is given by,

\begin{equation}
Q(x)=\frac{1}{1+\exp(-2\beta x)}
\label{prob}
\end{equation}

the stock state $s_{i}(t)$ is reinforcing or {\it persistent} so that
$s_{i}(t+1)=s_{i}(t)$, while with probability $1-Q(u_{i}(t))$ it
reverses or is {\it anti-persistent} so that
$s_{i}(t+1)=-s_{i}(t)$.

Therefore it is a basic assumption of this model that there are always
persistent and anti-persistent stocks and we can imagine how this
might occur as follows: Every time $t$, the speculators choose a pair
of stocks $\alpha$ and $\beta$ from, say, the excess demand stocks so
that $s_{\alpha}(t)=1$ and $s_{\beta}(t)=1$ and compare them by
comparing their confidences $V_{\alpha}(t)$ and
$V_{\beta}(t)$. Suppose $V_{\alpha}(t)>V_{\beta}(t)$, therefore the
speculators decide to demand $\alpha$ more and to supply $\beta$
instead. Therefore $s_{\alpha}(t+1)=1$ and
$s_{\beta}(t+1)=-1$. Conversely suppose the speculators are comparing
stocks $\alpha$ and $\beta$ which are in excess supply with
$s_{\alpha}(t)=-1$ and $s_{\beta}(t)=-1$, where
$V_{\alpha}(t)>V_{\beta}(t)$, then speculators decide to continue to
supply $\alpha$ while demanding $\beta$ so that $s_{\alpha}(t+1)=-1$
and $s_{\beta}(t+1)=1$. If we imagine this process of comparing stocks
happening continuously we can imagine each share generally interacting
with the market confidence $V(t)$ through the relative confidence
$u_{i}(t)=V_{i}(t)-V(t)$ and Eqn.\ref{prob} with the above $s_{i}(t)$
update dynamics becomes a plausible description. (In fact as an
alternative model, to be reported separately, we can say high valued
stocks $\alpha$ are persistent $s_{\alpha}(t+1)=s_{\alpha}(t)$ but
instead of low-valued stocks $\beta$ being anti-persistent we could
imagine that they are completely independent of the previous state
such that $s_{\beta}(t+1)=\pm1$ at random.) (We should also point out
that if the idea of anti-persistent reversing demand/supply stock
states seems strange so should the idea of persistent high volatility
itself!)

The relative confidence $u_{i}(t)$ measures to what extent speculators
have faith in the persistence of stock $i$'s state $s_{i}(t)$ compared
to their belief in the persistence of the whole market $V(t)$. The
`inverse temperature' parameter $\beta$ in Eq.\ref{prob} measures to
what extent speculator expectation is self-fullfilling, that is the
extent to which homogenous `herding' occurs. When $\beta=\infty$ the
system is deterministic and Eq.\ref{prob} reduces to a step function
and the state of stocks $i$, $s_{i}(t)$, with $V_{i}(t)>V(t)$ will be
persistent with probability 1, while stocks $i$ with $V_{i}(t)<V(t)$
will be anti-persistent with probability 1, i.e. complete homogenous
herding. In this case the speculators behave with `one mind'. With
$\beta=0$ all states $s_{i}(t+1)$ are chosen randomly independent of
$s_{i}(t)$ and speculators are therefore completely independent in
their viewpoints. (We can also say equivalently that $\beta$ measures
the extent to which traders {\it know} stock $i$'s confidence
$V_{i}(t)$.) It is a basic assumption of this model that a single
stock cannot be considered on it's own, independently of the rest of
market, speculators are always dynamically {\it choosing}. This is
reminiscent of an ecosystem where a single organism cannot be
considered independently of the other coevolving organisms in the
system.

\begin{figure}
\resizebox{0.45\textwidth}{8cm}{ 
\includegraphics{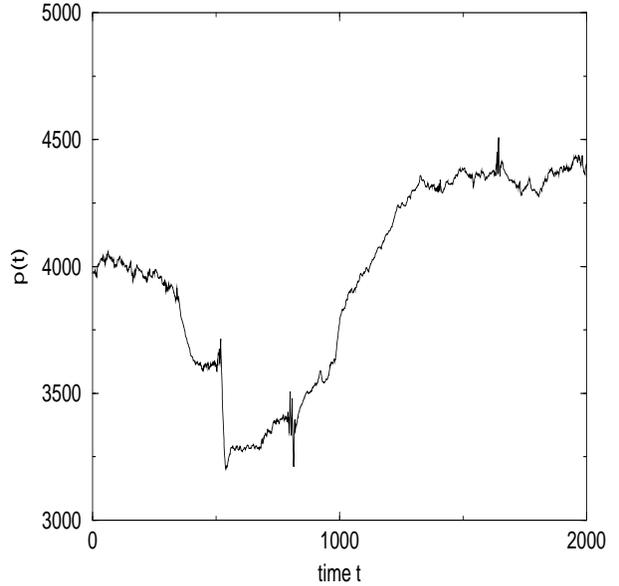}
}
\caption{Single realization of the market index price time series
$p(t)$ for $c=0.001$, $\beta=80$ and $N=200$.}
\label{fig:ts}       
\end{figure}

In the version of this model presented here, the dynamics of the
confidences $V_{i}(t)$ themselves are treated as internally defined
behaviour of the model. If a stock $i$ has persistent excess
demand/supply at time $t$ then we suggest speculators decrease their
confidence $V_{i}(t)$ in it by a small amount $c$, ($c>0$). If on the
other hand the excess demand/supply of stock $i$ is observed to
reverse we say speculator confidence is coupled to the excess
demand/supply state of the market $G(t)$ such that the change in
confidence, $\Delta V_{i}(t)=V_{i}(t+1)-V_{i}(t)$, is given by,

\begin{equation}
\Delta V_{i}(t)= \left \{\begin{array}{ll}-c,\hspace{1cm} & if
\hspace{1cm} s_{i}(t+1)=s_{i}(t)\\ s_{i}(t)G(t), & if\hspace{1cm}
s_{i}(t+1)=-s_{i}(t)
\end{array}
\right.
\label{values}
\end{equation}

Therefore since $s_{i}(t)G(t)=-\Delta s_{i}(t)G(t)/2$, speculator
confidence $V_{i}(t)$ in stock $i$ decreases when stock $i$'s state
$s_{i}(t)$ moves into the overall market majority state as measured by
$G(t)$. This means the probability of a stocks state reversing
increases when it moves into the majority and decreases when it moves
into the minority. This rule has some similarity to the Minority
Game\cite{MG}. We believe, however, that it is more appropriate for
financial market modeling to increase an agent's fitness when the
agent {\it moves into} the minority, rather than when it is {\it in}
the minority. As is well known, the way to make speculative money in
markets is to stay in the majority but switch state into the minority
just before everybody else does. (See also\cite{Farmer}.) Of course in
the Minority Game the agents are traders not stocks, but here we
assert that a similar effect occurs between individual stocks and the
market itself because of the effect of the coupling of the individual
stocks to the overall market. This coupling is due to the collective
choosing action of the background traders.

Our stock confidences $V_{i}(t)$ and the corresponding probabilities
$Q_{i}(t)$ are defined in cumulative terms, through continuous changes,
like cumulative fitnesses in ecodynamics.  We believe this is in
keeping with the way people think, traders do not simply forget their
previous evaluations of the stocks, constructing probabilities anew,
but rather dynamically update their perceptions, in a simple way
depending on the current behaviour of the stock.
 
While the parameter $\beta$ measures the amount of homogeneity of
speculator opinion in the market, $c$ measures the rate by which the
speculators in general lose faith in a current price trend
$s_{i}(t+1)=s_{i}(t)$. That is the rate by which they lose faith in a
stock's bullishness or bearishness.

The dynamic is synchronous, all $V_{i}(t)$ and $s_{i}(t)$ are updated
at the same $V(t)$ and $G(t)$ according to $Q(u_{i}(t))$. For initial
conditions the spins $s_{i}(0)$ are chosen randomly and the
confidences $V_{i}(0)$ randomly and uniformly on $[-1,1]$.

\section{Numerical Results. The Behaviour of the Volatility}
In our paper\cite{physA} we describe the behaviour of the model in
detail as we vary the parameters $c$ and $\beta$. As explained in that
paper we believe real financial markets are described by $c \approx 0$
($c>0$) and $\beta \approx 80$. We show in that paper that for those
parameters values the price returns distributions agree very well with
the empirical observations mentioned in the Introduction. A typical
index price time series for $c=0.001$, $\beta=80$ and $N=200$ is shown
in Fig.\ref{fig:ts}. This is just defined from the cumulative changes
as $p(t)=\sum_{j=1}^{t}\Delta p(j)$. Qualitatively it looks very
reminiscent of a real time series, with periods of time which look
like Gaussian random walks and other times which have larger
fluctuations. One `crash' is visible, as is a period of `sustained
growth'.

Here we study the behaviour of the volatility for the same parameter
values and compare it with real observations mentioned in the
Introduction. 

\begin{figure}
\resizebox{0.45\textwidth}{10cm}{ 
\includegraphics{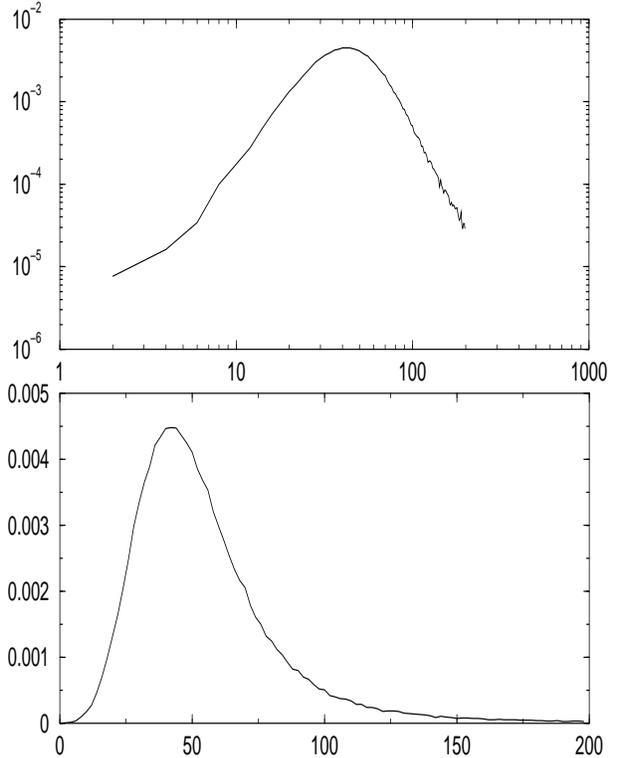}
}
\caption{Distribution of absolute price changes $|\Delta p(t)_{\Delta
t}|$ for $\Delta t=9$, $c=0.001$, $\beta=80$ and $N=200$. The
distribution is calculated from one single realization of the price
index time series. The lower panel shows the distribution in
linear-linear axes, the upper panel in log-log axes.}
\label{fig:vol}       
\end{figure}

Volatility is usually measured from financial market time series by
taking a certain time window and calculating the standard deviation of
the price fluctuations in that window. When considering time series of
length several years the window is often of the order of a few
weeks. Here, as a proxy for the volatility, we use the distribution of
the absolute changes in price in a similar way to\cite{stan3}. This is
shown in Fig.\ref{fig:vol} for the absolute values of $\Delta
p(t)_{\Delta t=9}=\sum_{j=t}^{t+9}\Delta p(j)$. We choose to show the
$\Delta t=9$ distribution, however other $\Delta t$ distributions are
similar. Shown in the bottom panel is the distribution in
linear-linear axes. The distribution is similar in appearance to the
log-normal distribution, although it has a long tail. The distribution
is shown in log-log axes in the upper panel of Fig.\ref{fig:vol}. The
tail is a straight line meaning the extreme absolute changes have a
power-law distribution. The slope of the tail is approximately 4, in
very good agreement with empirical results in \cite{stan3}, explained
in the Introduction.

\begin{figure}
\resizebox{0.45\textwidth}{8cm}{ 
\includegraphics{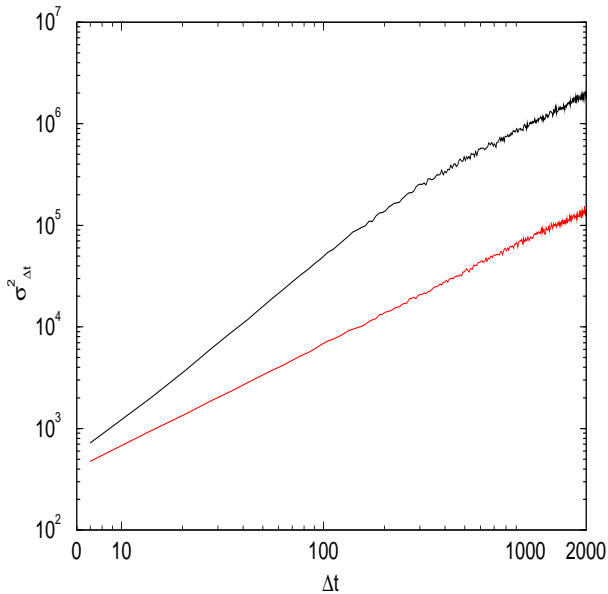}
}
\caption{Diffusion of the volatility for a single realization of the
price index time series with $c=0.001$, $\beta=80$ and $N=200$. The
variance $\sigma^{2}(\Delta t)$ is plotted against $\Delta t$ in
log-log axes. The upper curve is the results from the time series, the
lower curve is from the scrambled time series.}
\label{fig:diff}       
\end{figure}

Next we study the dynamical behaviour of the volatilty. In
\cite{physA} we show that the price changes time series from this
model show intermittent volatility clustering. In that paper we give
some explanation for this behaviour. For our chosen parameters
$c=0.001$ and $\beta=80$ the time series shows weak superdiffusion,
the characteristic of persistence. Shown in Fig.\ref{fig:diff} in
log-log axes is the variance $\sigma^{2}(\Delta t)$ of the price
changes $\Delta p(t)_{\Delta t}$ time series, plotted against $\Delta
t$. From the relationship,

\begin{equation}
\sigma^{2}(\Delta t)=(\Delta t)^{2\mu}
\end{equation}

we obtain $2\mu\approx 1.5$ for $\Delta t\approx 1 \sim 300$, while
for $\Delta t > 300$ we obtain $2\mu\approx 1.0$. This implies
superdiffusion at shorter time scales but with reversion to normal
diffusion for longer time scales. This is in very close agreement with
studies in \cite{stan3}, explained in the Introduction. Also shown in
Fig.\ref{fig:diff} is the same analysis for the scrambled time series,
performed as a check. As expected the slope reverts to that for normal
diffusion.

\section{Discussion}
We have described a multi-share financial market model and numerically
studied the behaviour of the volatility of the market index price
changes $\Delta p(t)$ for the parameter values $c=0.001$ and
$\beta=80$. Our results agree very well with empirical studies. In our
paper \cite{physA} we show these results are robust to changing the
value of $c$ providing it remains `small' or equivalently the system
size $N$ is large. Since $c$ measures the rate at which speculators
become nervous of continuing bullish or bearish trends, $c\to 0$,
($c>0$), means markets find a state where speculators try to
collectively `stand their ground' the longest. $\beta\approx 80$ in
fact corresponds to a phase transition between a regime where price
fluctuations obey Gaussian statistics and a regime where they obey
Levy distribution\cite{LP1} statistics. In \cite{physA} we give
qualitative reasons why markets should be found in this phase
transition region.


\end{document}